\documentclass[twoside,12pt]{article}
\usepackage{graphicx}

\newcommand{\be}{\begin{equation}}
\newcommand{\ee}{\end{equation}}
\newcommand{\bea}{\begin{eqnarray}}
\newcommand{\eea}{\end{eqnarray}}

\begin{document}

\title{
\parbox{.9\textwidth}{\flushright\large\rm \hfill KEK-TH-1508, RBRC-930}\\
Nucleon structure from 2+1-flavor dynamical DWF lattice QCD at nearly physical pion mass
}
\author{
Meifeng Lin$^{1,2}$ and Shigemi Ohta$^{3,4,2}$\\
for RBC and UKQCD Collaborations,\\
$^1$Department of Physics, Yale University, New Haven CT 06520, USA\\
$^2$RIKEN BNL Research Center, BNL, Upton, NY 11973, USA\\
$^3$Inst.\ of Particle and Nuclear Studies, KEK, Tsukuba, Ibaraki 3050801, Japan\\
$^4$SOKENDAI Graduate University, Hayama, Kanagawa 2400193, Japan
}
\date{Talk given at Erice School, 
September 16-24, 2011, Erice, Sicily}

\addtolength{\topmargin}{-5mm}
\addtolength{\textheight}{5mm}

\maketitle

\begin{abstract}
Domain-wall fermions (DWF) is a lattice discretization for Dirac fields that preserves continuum-like chiral and flavor symmetries that are essential in hadron physics.
RIKEN-BNL-Columbia (RBC) and UKQCD Collaborations have been generating sets of realistic 2+1-flavor dynamical lattice quantum chromodynamics (QCD) numerical ensembles with DWF quarks with strange mass set almost exactly at its physical value via reweighing and degenerate up and down mass set as light as practical.
In this report the current status of the nucleon-structure calculations using these ensembles are  summarized.
\end{abstract}


\section{Introduction}

Domain-wall fermions (DWF) \cite{Kaplan:1992bt,Shamir:1993zy,Furman:1994ky} is a lattice discretization scheme for Dirac fields that preserves continuum-like chiral and flavor symmetries.
The symmetries allow a straight-forward implementation of non-perturbative renormalization \cite{Dawson:1997ic} of electroweak transition matrix elements between hadronic states.
These are significant advantages over more conventional lattice discretizations that lack either or both of the symmetries, making even identification of hadronic states difficult, let alone the non-perturbative renormalization.
RIKEN-BNL-Columbia (RBC) Collaboration \cite{Blum:2000kn,Blum:2001xb,Aoki:2002vt,Aoki:2004ht,Aoki:2005ga} successfully demonstrated these advantages of DWF lattice quantum chromodynamics (QCD) using the QCDSP and QCDOC dedicated supercomputers \cite{QCDOCIBMJ} they designed and built.

More recently, RBC and UKQCD Collaborations have been generating sets of 2+1-flavor dynamical lattice quantum chromodynamics (QCD) numerical ensembles with DWF quarks where strange mass is set  almost exactly at its physical value and degenerate up and down mass as light as practical.
There have been three such sets \cite{Allton:2008pn,Aoki:2010dy,Jung:2010jt,Kelly:LAT2011} (see also Fig.\ \ref{fig:masssummary} where nucleon mass calculated in each of the sets is summarized):
\begin{figure}[tb]
\begin{center}
\includegraphics[width=.75\linewidth,clip]{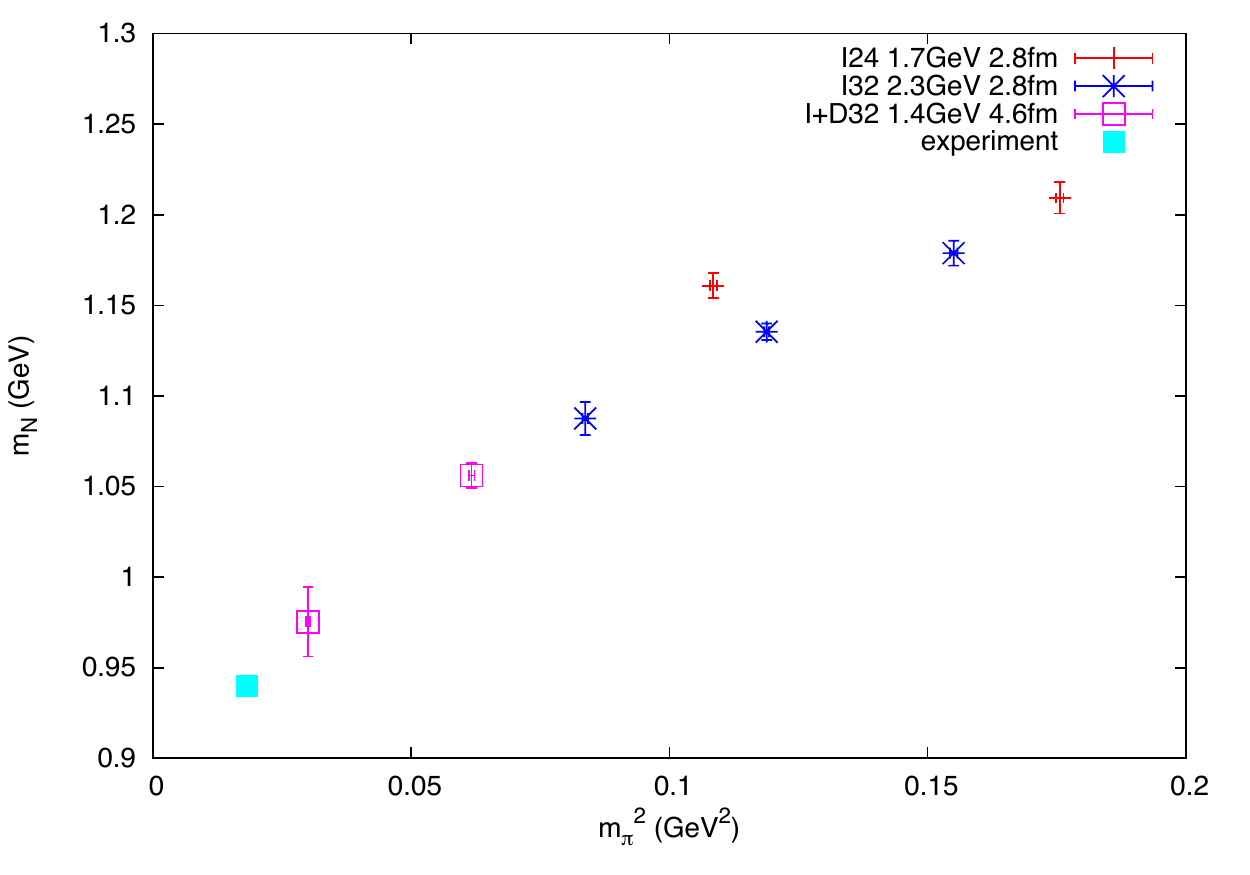}
\end{center}
\caption{
Nucleon mass from the RBC and UKQCD (2+1)-flavor dynamical DWF ensembles plotted against corresponding pion mass squared.  I24 and I32 are from ensembles with simple Iwasaki gauge action \cite{Allton:2008pn,Aoki:2010dy} while I+D32 are with new Iwasaki+DSDR gauge action \cite{Jung:2010jt,Kelly:LAT2011}.
}
\label{fig:masssummary}
\end{figure}
the first and second used Iwasaki gauge action at lattice cutoff, \(a^{-1}\), of about 1.7 GeV \cite{Allton:2008pn} and 2.2 GeV \cite{Aoki:2010dy} respectively, both with spatial volume of about \(({\rm 2.8 fm})^3\), and pion mass ranging from about 420 MeV to about 290 MeV.
We first use pion, kaon, and \(\Omega\)-baryon mass to set the physical quark mass and lattice scale and then obtain predictions for other observables such as pion and kaon decay constants with a few \% accuracy.
At this level of high accuracy our predictions are now not limited by the statistics but by poor applicability of chiral perturbation or other chiral extrapolation ansatz from the relatively heavy pion mass we used.

\addtolength{\topmargin}{5mm}
Thus the third set \cite{Jung:2010jt,Kelly:LAT2011} at lighter pion mass of about 250 MeV and 170 MeV are being produced.
This is made possible by using a new combination for gauge action of Iwasaki and a multiplicative dislocation-suppressing-determinant ratio (DSDR) factor \cite{Vranas:1999rz,Vranas:2006zk,
Renfrew:2009wu} that allows a lower cutoff of about 1.4 GeV while keeping the residual breaking of chiral symmetry sufficiently small.
The lower lattice cutoff also allows large spatial extent of about 4.6 fm which is important in studying larger hadrons such as nucleon.

\begin{figure}[tb]
\begin{center}
\includegraphics[width=.75\linewidth,clip]{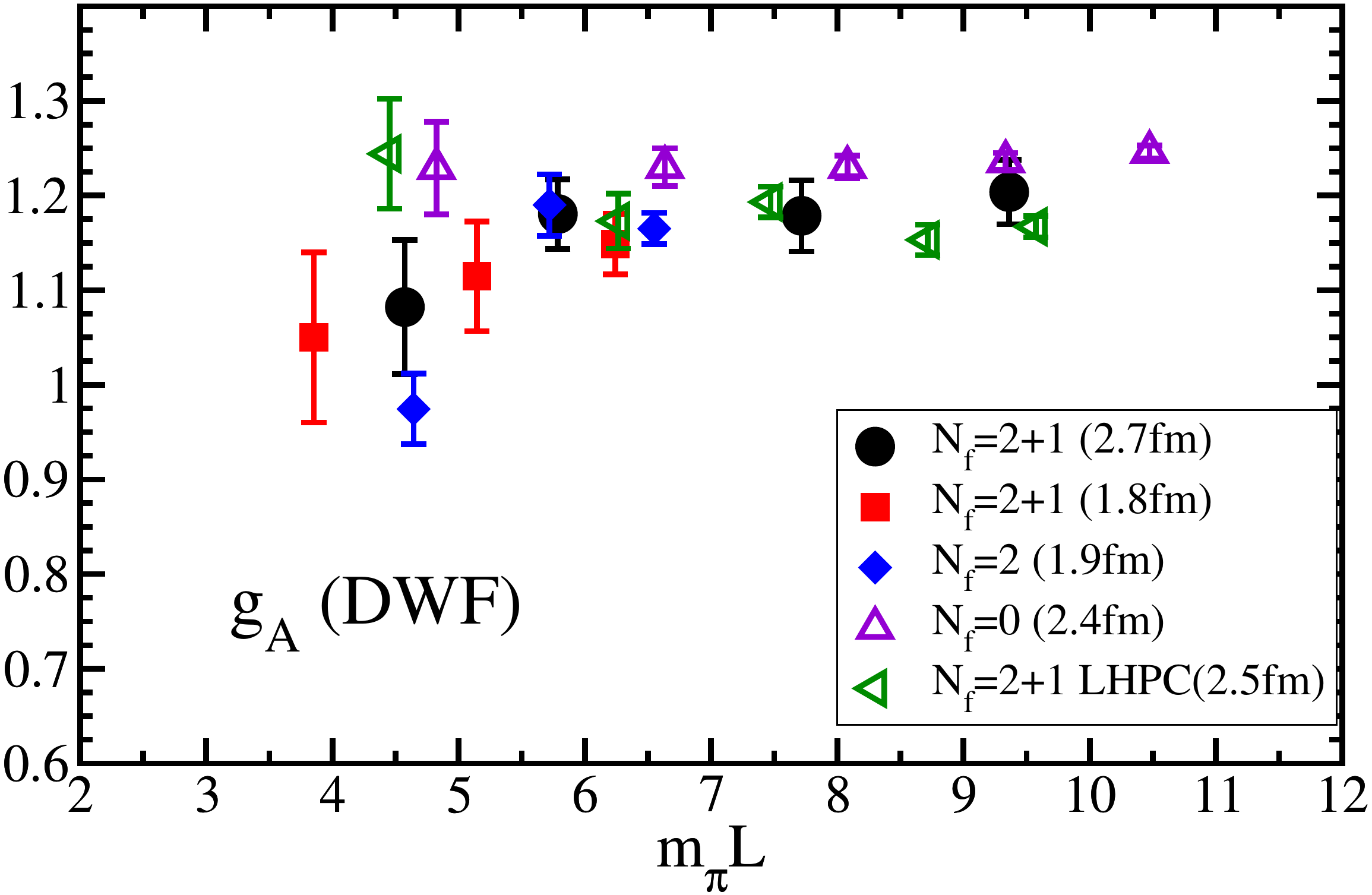}
\end{center}
\caption{Nucleon isovector axial charge, \(g_A\), calculated in numerical lattice QCD with DWF quarks plotted against a dimensionless variable, \(m_\pi L\), the product of calculate pion mass, \(m_\pi\), and lattice spatial extent, \(L\).
Results from RBC/UKQCD unitary (2+1)-flavor dynamical quark calculations show scaling in this variable with strong dependence.
In copmarison, non-unitary DWF calculations with either quenched or MILC ensembles show only weak, if at all, dependence.}
\label{fig:gAmpiL}
\end{figure}

We reported nucleon structure calculations using the first set of these gauge ensembles with 1.7-GeV cutoff and \(({\rm 2.8 fm})^3\) volume in Refs.\ \cite{Yamazaki:2008py,Yamazaki:2009zq,Aoki:2010xg}:
Most importantly we discovered a strong dependence on pion mass, \(m_\pi\), and lattice spatial extent, \(L\), in the isovector axial charge, \(g_A\), as summarized in Fig.\ \ref{fig:gAmpiL} \cite{Yamazaki:2008py}.
The dependence is much stronger than had been observed in non-unitary calculations with DWF quarks using either quenched \cite{Sasaki:2003jh} or rooted-staggered fermion \cite{Edwards:2005ym} ensembles.
As the dependence appears to scale with a dimensionless quantity, \(m_\pi L\), the product of calculated pion mass, \(m_\pi\), and the lattice spatial extent, \(L\), a likely explanation for the dependence is that significant part of the nucleon isovector axialvector current is carried by the expected, but never seen, ``pion cloud'' surrounding the nucleon.
If confirmed, this calculation may be the first concrete evidence of such a pion cloud.

Indeed similar strong dependence on pion mass and lattice spatial extent is seen in other axialvector-current form factors but not in the conserved vector-current ones \cite{Yamazaki:2009zq} or in low moments of the structure functions \cite{Aoki:2010xg}, supporting the pion-cloud interpretation.
However, since the calculations so far have only been carried out at single lattice cutoff, \(a^{-1}\), of about 1.7 GeV, two lattice spatial extent, \(L\), of about 1.8 and 2.7 fm, and relatively heavy pion mass down to 330 MeV, it seems premature to conclude that the dependence is caused by the pion cloud.On the other hand the findings were sufficient for us to decide skipping nucleon study using the second set of ensembles \cite{Aoki:2010dy} at lattice cutoff of about 2.2 GeV with even smaller \(m_\pi L\) at their lightest pion mass.

This possible discovery of pion cloud provides a good motivation for the new study being reported here with pion mass set at about 170 MeV and 250 MeV and lattice spatial extent of about 4.6 fm. The finite-size scaling parameter,  \(m_\pi L\), is respectively at 4.2 and 5.8 and should allow us to better understand the observed dependence.
In meson physics these ensembles help to improve our combined chiral and continuum limit study \cite{Jung:2010jt,Kelly:LAT2011}.
Such a study, however, requires good understanding of chiral and finite-size corrections to the observables which are unfortunately missing for baryons in general.
For this reason we are not yet attempting a similar combined chiral and continuum limit study for baryons at this moment.
Rather, we are refining our fixed-cutoff study of nucleon to improve our understanding of its chiral behavior.

\section{Nucleon structure at 1.7 GeV}

We here summarize the nucleon structure calculations \cite{Yamazaki:2008py,Yamazaki:2009zq,Aoki:2010xg} by RBC and UKQCD collborations using the 1.7-GeV dynamical (2+1)-DWF ensembles \cite{Allton:2008pn}.

In an earlier study with two dynamical DWF flavors \cite{Lin:2008uz}, we identified an important source of systematic error in lattice-QCD numerical calculation of nucleon structure, namely excited-state contamination.
As the amount of contamination varies depending on the shape of source smearing, we need to optimize the combination of source smearing and source-sink separation in order to filter out the excited-state contamination while maintaining reasonable statistical signal.
If we choose too long a separation then even the ground state decays and no signal is obtained.
For our choice of Gaussian source smearing width of 7 lattice units, a source-sink separation of about 12, or about 1.4 fm in physical unit, was optimal.

After making this important adjustment we discovered \cite{Yamazaki:2008py} the isovector axial charge strongly depends on pion mass and lattice size, as was discussed earlier with Fig.\ \ref{fig:gAmpiL}.
The dependence seems to scale in a single parameter, the product \(x=m_\pi L\) of the pion mass \(m_\pi\) and lattice size \(L\).
Though our dynamic range in the scaling parameter is rather narrow to distinguish various ansatze on functional form, \(f(x)\), of this scaling, and so cannot yet clarify if this is indeed the pion cloud, fitting to various forms such as \(x^{-3}\) (inverse volume) or \(e^{-x}/\sqrt{x}\) (pion cloud) is possible, and results in an estimate of \(g_A=1.20(6)_{\rm stat}(4)_{\rm syst}\).
This strong dependence on \(m_\pi L\) is observed also in other axialvector-current form factors \cite{Yamazaki:2009zq}.
In order to drive the systematic error arising from this dependence below 1 \%, we would need \(m_\pi L\) of 6 to 8, or \(L\) of about 5 fm for \(m_\pi\) of 300 MeV and 10 fm for physical pion.

In contrast to the axialvector-current form factors, the vector-current ones \cite{Yamazaki:2009zq} do not show dependence on the lattice size even at the lightest pion mass of about 330 MeV.
Mean squared radii of Dirac and Pauli form factors are obtained 
without chirally expected singular behavior in \(m_\pi^2\), and undershoot the experiments.
On the other hand the anomalous magnetic moment is in rough agreement with the experiment.These observations are confirmed by a LHP study \cite{Lin:2010ne} using our 2.2-GeV ensembles \cite{Aoki:2010dy}.

In low moments of isovector structure functions \cite{Aoki:2010xg}, such as quark momentum fraction, \(\langle x \rangle_{u-d}\), or helicity fraction, \(\langle x \rangle_{\Delta u - \Delta d}\), no dependence on \(m_\pi L\) is seen either .
In a naturally renormalized ratio, \(\langle x \rangle_{u-d}/\langle x \rangle_{\Delta u - \Delta d}\), of the two fractions, no dependence on the mass, \(m_q \propto m_\pi^2\), is seen either and the agreement with experiment is excellent.
No dependence on \(L\) is seen in respective fractions either.
Both show interesting trending down toward the experiment, motivating us to calculate these quantities at lower pion mass.

\section{Status at 1.4 GeV}

\begin{figure}[tb]
\begin{center}
\includegraphics[width=.75\linewidth,clip]{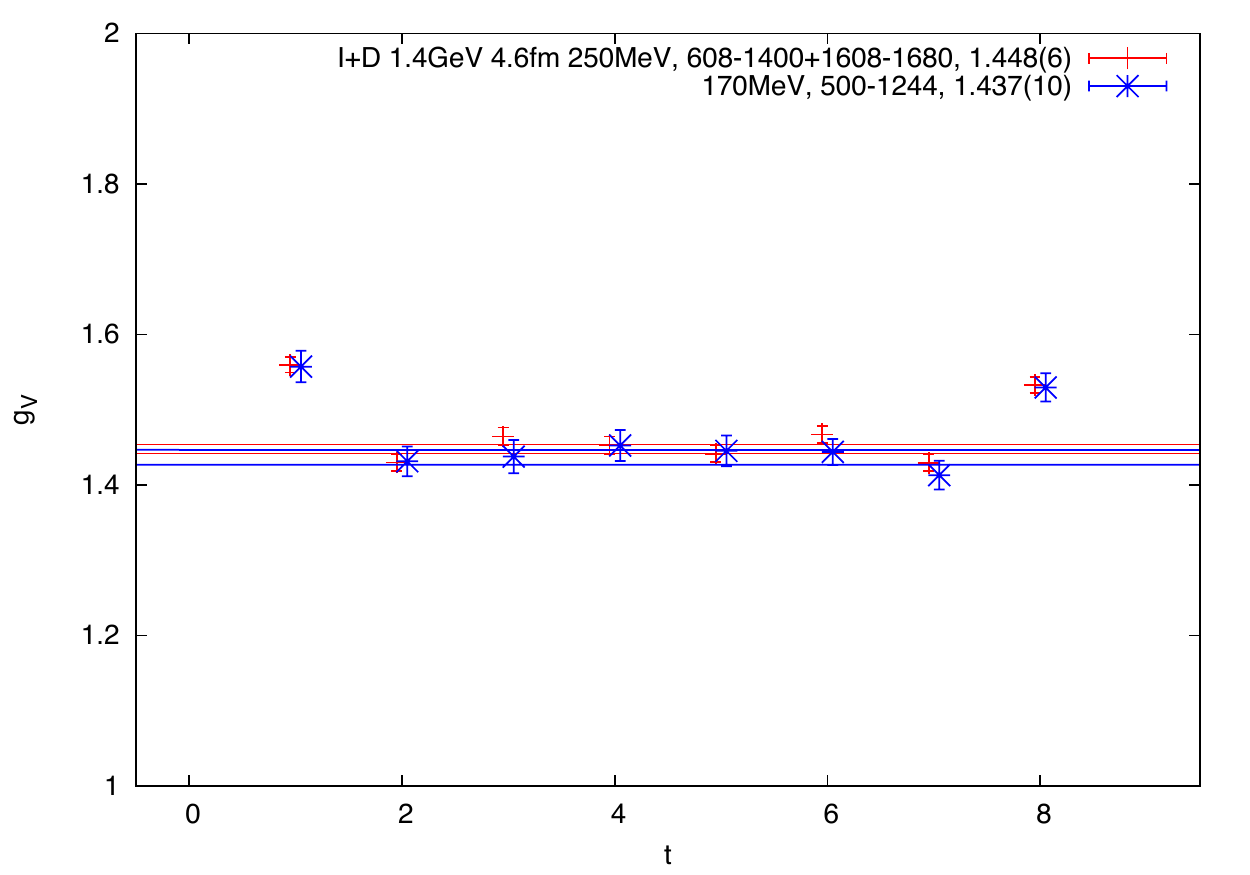}
\end{center}
\caption{
Clear signals are obtained for local-current isovector vector charge of nucleon, yielding values \(g_V = 1.447(6)\) and 1.437(10), for 250-MeV and 170-MeV ensembles, respectively.
From their inverses a vector current renormalization of \(Z_V = 0.700(9)\) in the chiral limit is obtained, in agreement with \(Z_A\) obtained in the meson sector, proving good chiral symmetry of the calculation.
}
\label{fig:gV}
\end{figure}
Our studies \cite{Yamazaki:2008py,Yamazaki:2009zq,Aoki:2010xg} of nucleon structure summarized in the previous section using the  (2+1)-flavor dynamical DWF lattice QCD at lattice cutoff of about 1.7 GeV in \((\sim{\rm 3 fm})^3\) spatial box and down to about 330 MeV pion mass \cite{Allton:2008pn} clearly point to the need of numerical calculations at yet lighter pion mass,  for better understanding of chiral behavior in the baryon sector.
Such need for lighter pion mass also arose from our continuum physics study in the meson sector combining the 1.7-GeV and 2.2-GeV ensembles \cite{Aoki:2010dy}.
Thus RBC and UKQCD collaborations started to generate a new set of ensembles with pion mass at about 170 and 250 MeV \cite{Jung:2010jt,Kelly:LAT2011}.
As the lower pion mass demands larger lattice spatial extent, the new ensembles are generated with about 4.6 fm spatial extent that translates to the scaling parameter \(m_\pi L\) of above 4 for the lighter mass and almost 6 for the heavier.
This is made possible by a newly developed gauge action with a multiplicative dislocation-suppressing-determinant-ratio (DSDR) factor \cite{Vranas:1999rz,Vranas:2006zk,
Renfrew:2009wu}.
We have about 2000 hybrid Monte Carlo time units for each ensemble, of which first 600 and 500 respectively for the heavy and light ensembles are discarded for thermalization.
We analyze every eight time unit with four evenly separated nucleon sources each.

We use Gaussian smearing \cite{Alexandrou:1992ti, Berruto:2005hg} for nucleon source to optimize the overlap with the ground state and compared the cases for widths 4 and 6 lattice units.
We found width 6 is better for both pion mass \cite{ohta:2010sr}.
We quote preliminary nucleon mass estimates of 0.721(13) and 0.763(10) lattice units which correspond to about 0.98 and 1.05 GeV with another preliminary estimate for the lattice cutoff of 1.368(7) GeV.
As are summarized in Fig.\ \ref{fig:masssummary}, these ensembles are nicely filling the gap toward the physical point.

Nucleon isovector vector-current and axialvector-current form factors and some low moments of isovector structure functions are being calculated using the RIKEN Integrated Cluster of Clusters, RIKEN, Wako, Japan for the 250-MeV ensemble and Lonestar and other clusters of US NSF Teragrid/XSEDE for the 170-MeV one.
By the time of this Erice school, 89 configurations for the former and the 99 for the latter had been analyzed.
Since then we have increased the statistics to 110 for the former and 94 for the latter, resulting in improved analyses:

The signals for the isovector vector charge, \(g_V\), are summarized in Fig.\ \ref{fig:gV}.
Estimates of  \(g_V = 1.448(6)\) and 1.437(10) are obtained for 250-MeV and 170-MeV ensembles, respectively.
The values deviate from unity because we use the local-current that are proportional to the conserved vector current.
From their inverses the renormalization of \(Z_V = 0.700(9)\) in the chiral limit is obtained for this local isovector vector current.
Considering the expected \(O(a^2)\) discretization error, this is in good agreement with \(Z_A = 0.6871(4)\) \cite{Kelly:LAT2011} for the local axial vector current obtained in the meson sector, and proves good chiral symmetry of the calculation.

\begin{figure}[tb]
\begin{center}
\includegraphics[width=.75\linewidth,clip]{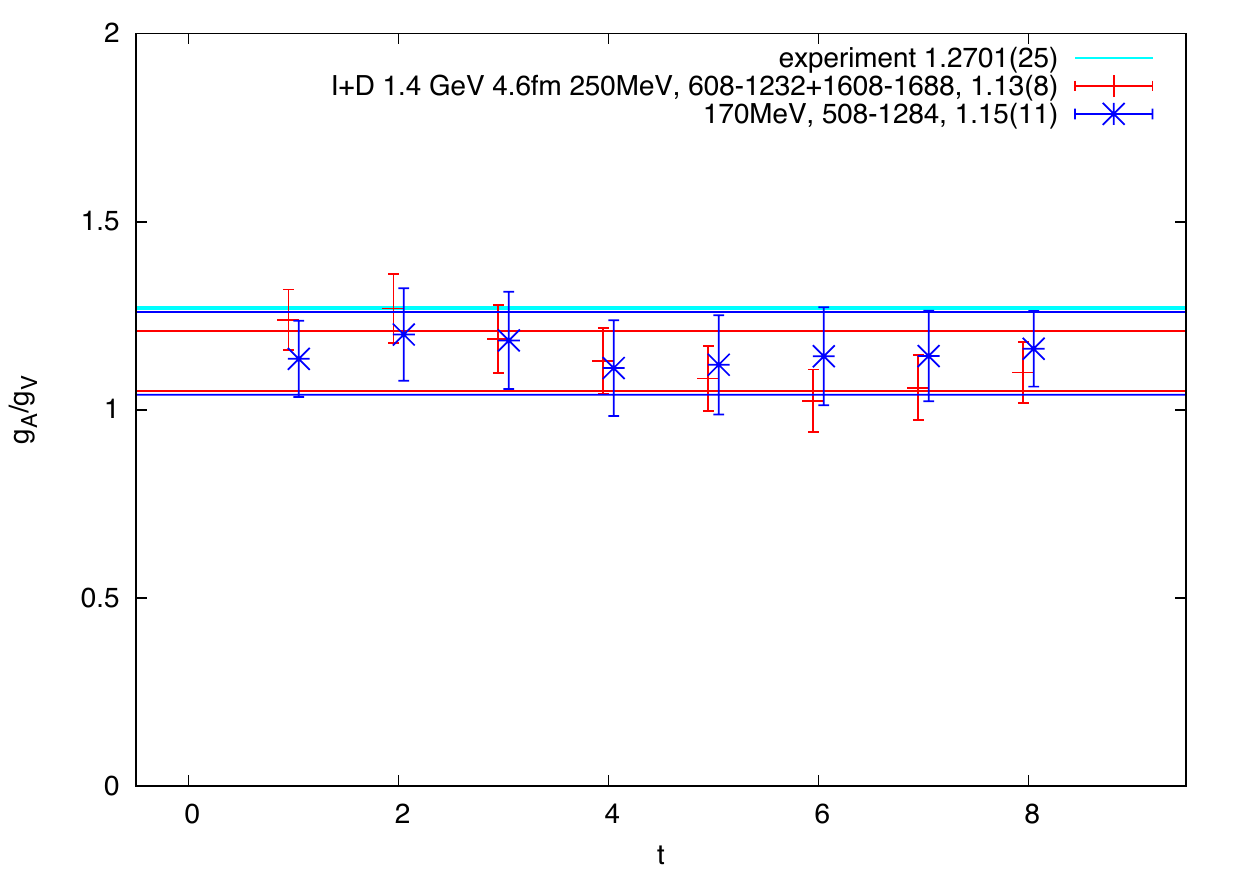}
\end{center}
\caption{
Signals for the ratio, \(g_A/g_V\), of the isovector vector charge, \(g_V\), and axial charge, \(g_A\).
While signals for the latter alone is much noisier than for the former, this ratio is less noisy, as the two share common non-perturbatice renormalization up to \(O(a^2)\) error because of good chiral symmetry of DWF discretization, and hence is directly comparable with the experiment.
With a fitting range from \(t=2\) to 7 we obtain estimates of 1.13(8) for the 250-MeV and 1.15(11) for the 170-MeV ensembles.
}
\label{fig:gAgV}
\end{figure}

The signals for the axial charge, \(g_A\), are much noisier.
However, as expected, the signals for the ratio of the axial and vector charges, \(g_A/g_V\), are less noisy, and are summarized in Fig.\ \ref{fig:gAgV}.
By fitting the plateaux from \(t=2\) to 7, we obtain estimates of \(g_A/g_V = 1.13(8)\) for the heavy and 1.15(11) for the light ensembles respectively.
The dependence on pion mass squared, \(m_\pi^2\), with the statistical error still at about ten percent, is difficult to judge.
The dependence on the finite-size scaling parameter, \(m_\pi L\), is presented in Fig.\ \ref{fig:gAgVmpiL}:
\begin{figure}[tb]
\begin{center}
\includegraphics[width=.75\linewidth,clip]{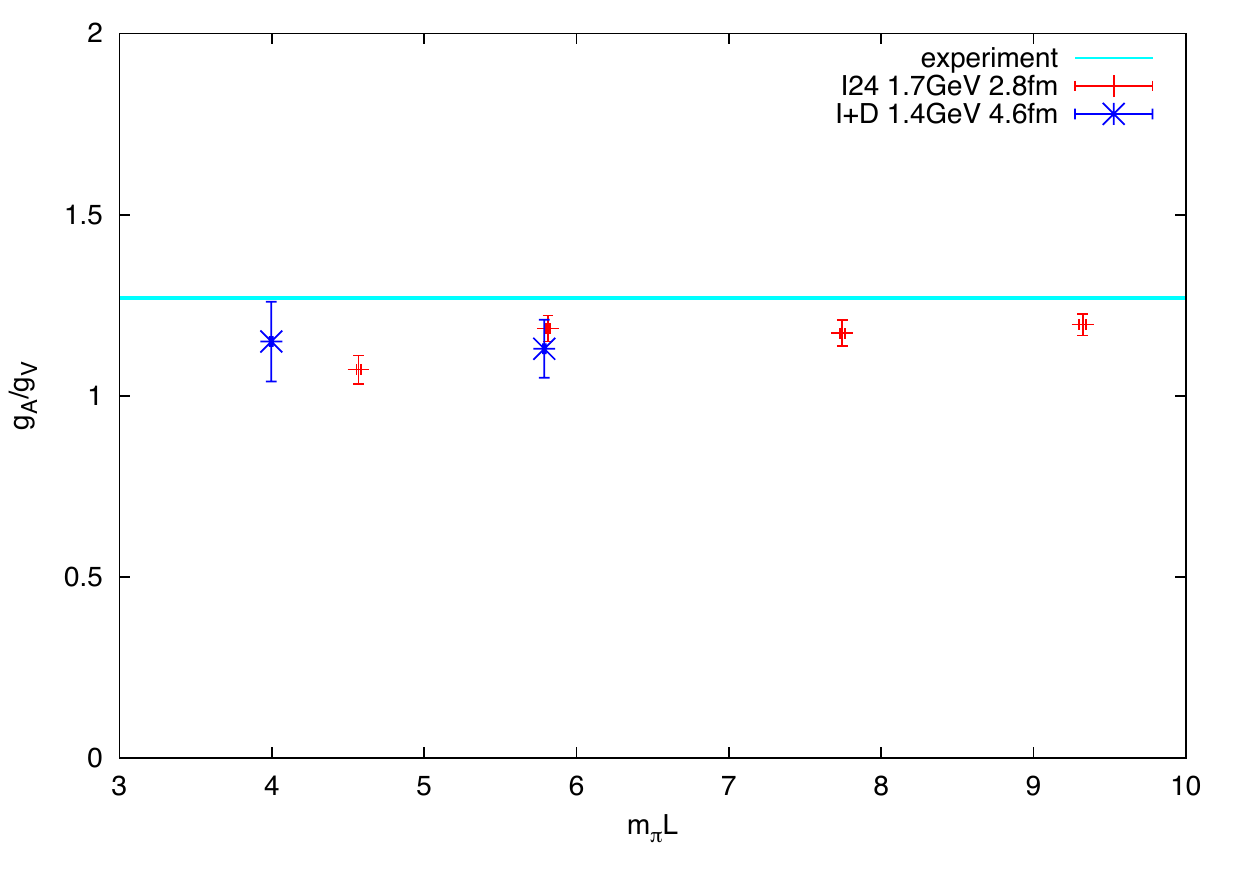}
\end{center}
\caption{
Dependence of the ratio, \(g_A/g_V\), on the finite-size scaling parameter, \(m_\pi L\), the product of calculated pion mass, \(m_\pi\), and linear spatial extent of the lattice, \(L\), for our earlier 1.7-GeV (red) and present 1.4-GeV (blue) studies.
Albeit with low statistics, the present results are consistent with the scaling in \(m_\pi L\) observed earlier.
}
\label{fig:gAgVmpiL}
\end{figure}
The heavy ensemble result is consistent with the finite-size effect seen at similar \(m_\pi L \sim 6\) in the 1.7-GeV study.
At a lighter pion mass and with a smaller $m_\pi L$, the result for the light ensemble roughly agrees with the finite-size scaling behavior discovered earlier.
However, the still-large statistical errors prevent us from drawing any definitive conclusions.

We also calculate the quark isovector momentum, \(\langle x \rangle_{u-d}\), and helicity, \(\langle x \rangle_{\Delta u - \Delta d}\), fractions and their ratio.
They are still rather noisy at the level of statistics, but seem to confirm that they also share common renormalization.
The ratio, albeit with large statistical error, seem consistent with the experiment.
The individual fractions seem to trend down with decreasing mass as well.

We thank RBC and UKQCD Collaborations, especially Yasumichi Aoki, Tom Blum, Chris Dawson, Taku Izubuchi, Chulwoo Jung, Shoichi Sasaki and Takeshi Yamazaki.
RIKEN, BNL, the U.S.\ DOE, University of Edinburgh, and the U.K.\  PPARC provided  facilities essential for the completion of this work.
The I+DSDR ensembles are being generated at ANL Leadership Class Facility (ALCF.)
The nucleon two- and three-point correlators are being calculated at RIKEN Integrated Cluster of Clusters (RICC) and US NSF Teragrid/XSEDE clusters.

\end{document}